# Flash-Based Extended Cache for Higher Throughput and Faster Recovery

Woon-Hak Kang[‡]
woonagi319@skku.edu

Sang-Won Lee[‡]
swlee@skku.edu

Bongki Moon[§]
bkmoon@cs.arizona.edu

[‡]College of Info. & Comm. Engr., Sungkyunkwan University, Suwon 440-746, Korea
[§]Dept. of Computer Science, University of Arizona, Tucson, Arizona, 85721, USA

## ABSTRACT

Considering the current price gap between disk and flash memory drives, for applications dealing with large scale data, it will be economically more sensible to use flash memory drives to supplement disk drives rather than to replace them. This paper presents *FaCE*, which is a new low-overhead caching strategy that uses flash memory as an extension to the DRAM buffer. *FaCE* aims at improving the transaction throughput as well as shortening the recovery time from a system failure. To achieve the goals, we propose two novel algorithms for flash cache management, namely, *Multi-Version FIFO replacement* and *Group Second Chance*. One striking result from *FaCE* is that using a small flash memory drive as a caching device could deliver even higher throughput than using a large flash memory drive to store the entire database tables. This was possible due to *flash write optimization* as well as *disk access reduction* obtained by the *FaCE* caching methods. In addition, *FaCE* takes advantage of the non-volatility of flash memory to fully support database recovery by extending the scope of a persistent database to include the data pages stored in the flash cache. We have implemented *FaCE* in the PostgreSQL open source database server and demonstrated its effectiveness for TPC-C benchmarks.

## 1. INTRODUCTION

As the technology of flash memory solid state drives (SSD) continues to advance, they are increasingly adopted in a wide spectrum of storage systems to deliver higher throughput at more affordable prices. For example, it has been shown that flash memory SSDs can outperform disk drives in throughput, energy consumption and cost effectiveness for database workloads [12, 13]. Nevertheless, it is still true that the price per unit capacity of flash memory SSDs is higher than that of disk drives, and the market trend is likely to continue for the foreseeable future. Therefore, for applications dealing with large scale data, it may be economically more sensible to use flash memory SSDs to supplement disk drives rather than to replace them.

In this paper, we present a low-overhead strategy for using flash memory as an extended cache for a recoverable database. This method is referred to as *Flash as Cache Extension* or *FaCE* for short. Like any other caching mechanism, the objective of *FaCE* is to provide *the performance of flash memory* and *the capacity of disk* at as little cost as possible. We set out to achieve this with the realization that flash memory has drastically different characteristics (such as no-overwriting, slow writes, and non-volatility) than DRAM buffer and they must be dealt with carefully and exploited effectively. There are a few existing approaches that store frequently accessed data in flash memory drives by using them as either faster disk drives or an extension to the RAM buffer. The *FaCE* method we propose is in line with the existing cache extension approaches but it is also different from them in many ways.

First, *FaCE* aims at *flash write optimization* as well as *disk access reduction*. Unlike a DRAM buffer that yields uniform performance for both random and sequential accesses, the performance of flash memory varies considerably depending on the type of operations (*i.e.*, read or write) and the pattern of accesses (*i.e.*, random or sequential). With most contemporary flash memory SSDs, random writes are slower than sequential writes approximately by an order of magnitude. *FaCE* provides flash-aware strategies for managing the flash cache that can be designed and implemented independently of the DRAM buffer management policy. By turning small random writes to large sequential ones, high sequential bandwidth and internal parallelism of modern flash memory SSDs can be utilized more effectively for higher throughput [5].

One surprising consequence we observed was that a disk-based OLTP system with a small flash cache added outperformed, by almost a factor of two, even an OLTP system that stored the entire database on flash memory devices. This result, which has never been reported to the best of our knowledge, demonstrates that the *FaCE* method provides a cost-effective performance boost for a disk-based OLTP system.

Second, *FaCE* takes advantage of the non-volatility of flash memory and extends the scope of a persistent database to include the data pages stored in the flash cache. Once a data page evicted from the DRAM buffer is staged in the flash cache, it is considered having been propagated to the persistent database. Therefore, the data pages in the flash cache can be utilized to minimize the recovery overhead, accelerate restarting the system from a failure and achieve transaction atomicity and durability at the nominal cost. The only additional processing required for a system restart is to restore the metadata of flash cache, but it does not take more than just a few seconds. In addition, since most data pages that need to be accessed during the recovery phase can be found in the flash cache, the recovery time will be significantly shortened. The recovery manager of *FaCE* provides mechanisms that fully support database recovery and persistent metadata management for cached pages.

Third, *FaCE* is a low-overhead framework that uses a flash memory drive as an extension of a DRAM buffer rather than a disk replacement. The use of flash cache is tightly coupled with the DRAM buffer. Unlike some existing approaches, a data page is cached in flash memory *not on entry* to the DRAM buffer, but instead *on exit* from it. This is because a copy in the flash cache will never be accessed while another copy of the same page exists in the DRAM buffer. The run-time overhead is very little, as there is no need for manual or algorithmic elaboration to separate hot data items from the cold ones. *FaCE* decreases but never increases the amount of traffic to and from disk, because it does not migrate data items between flash memory and disk drives just for the sake of higher cache hits. When a page is to be removed from the flash cache, if it is valid and dirty, it will be written to disk, very much like a dirty page evicted from the DRAM buffer and flushed to disk.

We have implemented the *FaCE* system in the PostgreSQL open source database server. Most of the changes necessary for *FaCE* are made within the buffer manager module, the checkpoint process and the recovery daemon. In the TPC-C benchmarks, we observed that *FaCE* yielded hit rates from the flash cache in the range from low 60 to mid 80 percent and reduced disk writes 50 to 70 percent consistently. This high hit rate and considerable write reduction led to substantial improvement in transaction throughput by up to a factor of two or more. Besides, *FaCE* reduced the restart time by more than a factor of four consistently across various checkpoint intervals.

The rest of the paper is organized as follows. Section 2 reviews the previous work on flash memory caching and presents the analysis of the cost effectiveness of flash memory as an extended cache. In Section 3, we overview the basic framework and design alternatives of *FaCE* and present the key algorithms for flash cache management. Section 4 presents the database recovery system designed for *FaCE*. In Section 5, we analyze the performance impact of the *FaCE* system on the TPC-C workloads and demonstrate its advantages over the existing approaches. Lastly, Section 6 summarizes the contributions of this paper.

## 2. BACKGROUND

As the technology of flash memory SSDs continues to innovate, the performance has improved remarkably in both sequential bandwidth and random throughput. Table 1 compares the price and performance characteristics of flash memory SSDs and magnetic disk drives. The throughput and bandwidth in the table were measured by Orion calibration tool [15] when the devices were in the steady state. The flash memory SSDs are based on either MLC or SLC type NAND flash chips. The magnetic disks are enterprise class 15k RPM drives, and they were tested as a single unit or as a 8-disk RAID-0 array.

### 2.1 Faster Disk vs. Buffer Extension

In some of the existing approaches [3, 11], flash memory drives are used as yet another type of disk drives with faster access speed. Flash memory drives may replace disk drives altogether for small to medium scale databases or may be used more economically to store only frequently accessed data items for large scale databases. When both types of media are deployed at the same time, a data item typically resides exclusively in either of the media unless disk drives are used in a lower tier in the storage hierarchy [8].

One technical concern about this approach is the cost of identifying hot data items. This can be done either statically by profiling [3] or dynamically by monitoring at run-time [11], but not without drawbacks. While the static approach may not cope with changes in access patterns, the dynamic one may suffer from excessive run-time and space overheads for identifying hot and data objects and migrating them between flash memory and disk drives. It is shown that the dynamic approach becomes less effective when the workload is update intensive [3].

In contrast, if flash memory SSDs are used as a DRAM buffer extension or a cache layer between DRAM and disk, it can simply go along with the data page replacement mechanism provided by the DRAM buffer pool without having to provide any additional mechanism to separate hot data pages from cold ones. There is no need to monitor data access patterns, hence very little run-time overhead and no negative impact from the prediction quality of future data access patterns.

Two important observations can be made in Table 1 in regard to utilizing flash memory as a cache extension. First, there still exists considerable bandwidth disparity between random writes and sequential writes with both SLC-based and MLC-based flash memory devices. The random write bandwidth was in the range of merely 10 to 13 percent of sequential write bandwidth, while random read enjoys bandwidth much closer (in the range of 48 to 60 percent) to that of sequential read. Unlike a DRAM buffer that yields uniform performance regardless of types or patterns of accesses, the design of flash cache management should take the unique characteristics of flash memory into account.

Second, disk arrays are a very cost-effective means for providing high sequential bandwidth. However, they still fall far behind flash memory SSDs with respect to random I/O throughput, especially for random reads. Therefore, the caching framework for flash memory should be designed such that random disk I/O operations are replaced by random flash read and sequential flash write operations as much as possible.

### 2.2 Cost-Effectiveness of Flash Cache

With a buffer replacement algorithm that does not suffer from Belady's anomaly [1], an increase in the number of buffer frames is generally expected to provide a fewer page faults. Tsuei *et al.* have shown that the data hit rate is a linear function of $\log(BufferSize)$ when the database size is fixed [18]. Based on this observation, we analyze the cost-effectiveness of flash memory as a cache extension. The question we are interested in answering is *how much flash memory will be required to achieve the same level of reduction in I/O time obtained by an increase in DRAM buffer capacity*. In the following analysis, $C_{disk}$ and $C_{flash}$ denote the time taken to access a disk page and the time

| Storage | 4KB Random Throughput (IOPS) | | Sequential Bandwidth (MB/sec) | | Capacity | Price in $ |
| Media | Read | Write | Read | Write | in GB | ($/GB) |
| --- | --- | --- | --- | --- | --- | --- |
| MLC SSD[†] | 28,495 | 6,314 | 251.33 | 242.80 | 256 | 450 (1.78) |
| MLC SSD[‡] | 35,601 | 2,547 | 258.70 | 80.81 | 80 | 180 (2.25) |
| SLC SSD[§] | 38,427 | 5,057 | 259.2 | 195.25 | 32 | 440 (13.75) |
| Single disk[¶] | 409 | 343 | 156 | 154 | 146.8 | 240 (1.63) |
| 8-disk[¶] RAID-0 | 2,598 | 2,502 | 848 | 843 | 1,170 | 1,920 (1.63) |

SSD: [†]Samsung 470 Series 256GB, [‡]Intel X25-M G2 80GB, [§]Intel X25-E 32GB
[¶]Disk: Seagate Cheetah 15K.6 146.8GB

Table 1: Price and Performance Characteristics of Flash Memory SSDs and Magnetic Disk Drives

taken to access a flash page, respectively.

Suppose the DRAM buffer size is increased from $B$ to $(1+\delta)B$ for some $\delta > 0$. Then the increment in the hit rate is expected to be

$$\alpha \log((1+\delta)B) - \alpha \log(B) = \alpha \log(1+\delta)$$

for a positive constant $\alpha$. The increased hit rate will lead to reduction in disk accesses, which accounts for reduced I/O time by $\alpha C_{disk} \log(1+\delta)$. If the $\delta B$ increment of DRAM buffer capacity is replaced by an extended cache of $\theta B$ flash memory, then the data hit rate (for both DRAM hits and flash memory hits) will increase to $\alpha \log((1+\theta)B)$. Since the rate of DRAM hits will remain the same, the rate of flash memory hits will be given by

$$\alpha \log((1+\theta)B) - \alpha \log(B) = \alpha \log(1+\theta).$$

Each flash memory hit translates to *an access to disk replaced by an access to flash memory*. Therefore, the amount of reduced I/O time by the flash cache will be $\alpha(C_{disk} - C_{flash}) \log(1+\theta)$.

The break-even point for $\theta$ is obtained by the following equation

$$\alpha C_{disk} \log(1+\delta) = \alpha(C_{disk} - C_{flash}) \log(1+\theta)$$

and is represented by a formula below.

$$1 + \theta = (1+\delta)^{\frac{C_{disk}}{C_{disk} - C_{flash}}}$$

For most contemporary disk drives and flash memory SSDs, the value of $\frac{C_{disk}}{C_{disk} - C_{flash}}$ is very close to one. For example, with a Seagate disk drive and a Samsung flash memory SSD shown in Table 1, the value of the fraction is approximately 1.006 for read-only workload and 1.025 for write-only workload.

This implies that the effect of disk access reduction by flash cache extension is almost as good as that of extended DRAM cache. Given that NAND flash memory is almost ten times cheaper than DRAM with respect to price per capacity and the price gap is expected to grow, this analysis demonstrates that the cost-effectiveness of flash cache is indeed significant.

### 2.3 Related Work

Flash memory SSDs have recently been adopted by commercial database systems to store frequently accessed data pages. For example, Oracle Exadata caches hot data pages in flash memory when they are fetched from disk [16, 17]. Hot data selection is done statically by the types of data such that tables and indexes have higher priority than log and backup data.

Similarly, the Bufferpool Extension prototype of IBM DB2 proposes a temperature-aware caching (TAC) scheme that relies on data access frequencies [2, 4]. TAC monitors data access patterns continuously to identify hot data regions at the granularity of an extent or a fixed number of data pages. Hot data pages are cached in the flash memory cache *on entry* to the DRAM buffer from disk. TAC adopts a write-through caching policy. When a dirty page is evicted from the DRAM buffer, it is written to both the flash cache and disk. Consequently, the flash cache provides caching effect for read operations but no effect of reducing disk write operations. Besides, the high cost of maintaining cache metadata persistently in flash memory degrades the overall performance. (See Section 4.1 for more detailed descriptions of the persistent metadata management by TAC.)

In contrast, the Lazy Cleaning (LC) method presented in a recent study caches data pages in flash memory *upon exit* from the DRAM buffer and handles them by a write-back policy if they are dirty [6]. It uses a background thread to flush dirty pages from the flash cache to disk, when the percentage of dirty pages goes beyond a tunable threshold. The LC method manages the flash cache using LRU-2 replacement algorithm. Hence replacing a page in the flash cache incurs costly random read and random write operations. Besides, since no mechanism is provided for incorporating the flash cache in database recovery, dirty pages cached in flash memory are subject to checkpointing to disk. It has been reported in the study that the additional cost of checkpointing is significant [6].

Among the aforementioned flash caching approaches, the LC method is the closest to the *FaCE* caching scheme presented in this paper. In both approaches, data pages are cached in flash memory upon exit from the DRAM buffer and managed by a write-back policy. Other than those similarities, the *FaCE* method differs from the LC method in a few critical aspects of utilizing flash memory for caching and database recovery purposes, and achieves almost twice higher transaction throughput than the LC method.

First of all, *FaCE* considers not only cache hit rates but also *write optimization* of flash cache. It manages the flash cache in the first-in-first-out fashion and allows multiple versions left in the flash cache, so that random writes are avoided or turned into sequential ones. The FIFO style replacement allows *FaCE* to replace a group of pages at once so that internal parallelism of a flash memory SSD can be exploited. It also boosts the cache hit rate by allowing a second chance for warm data pages to stay in the flash cache. In general, a flash caching mechanism improves the throughput of an OLTP system as the capacity of its flash cache increases, which will peak when the entire database is stored in the flash cache. With the *FaCE* method, however, we observed that a disk-based OLTP system achieved almost twice higher transaction throughput than an OLTP system based entirely on flash memory, by utilizing a flash cache whose size is only a small fraction (about 10%) of the database size.[1]

Second, unlike the LC method, *FaCE* extends the scope of a persistent database to include the data pages stored in the flash cache. It may sound a simple notion but its implications are significant. For example, database checkpointing can be done much more efficiently by flushing dirty pages to the flash cache rather than disk and by not subjecting data pages in the flash cache to checkpointing. Furthermore, the persistent data copies stored in the flash cache can be utilized for faster database recovery from a failure. *FaCE* provides a low-overhead mechanism for maintaining the metadata persistently in flash memory. We have implemented all the caching and recovery mechanisms of *FaCE* fully in the PostgreSQL system.

## 3. FLASH AS CACHE EXTENSION

The principal benefit of using flash memory as an extension to a DRAM buffer is that a mix of flash memory and disk drives can be used in a unified way without manual or algorithmic intervention for data placement across the drives. When a data page is deemed cold and swapped out by the DRAM buffer manager, it may be granted a chance to stay in the flash cache for an extended period. If the data page turns out warm enough to be referenced again while staying in the flash cache, then it will be swapped back into the DRAM buffer from flash memory much more quickly than from disk. If a certain page is indeed cold and not referenced for a long while, then it will not be cached in either the DRAM buffer or the flash cache.

### 3.1 Basic Framework

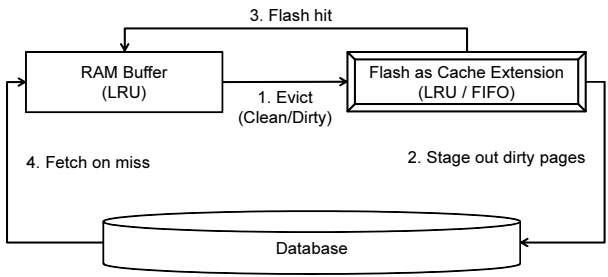

Figure 1: Flash as Cache Extension: Overview

---

[1]Performance gain by the LC method was measured when the flash cache was larger than half of the entire database [6]. Therefore, the level of transaction throughput reported in the study is expected to fairly close to what would be achieved by storing the entire database in flash memory devices.

Figure 3.1 illustrates the main components and their interactions of a caching system deployed for a database system, when the flash cache is enabled. The main interactions between the components are summarized as follows.

- When the database server requests a data page that is not in memory (*i.e.*, a DRAM buffer miss), the flash cache is searched for the page. (A list of pages cached in flash memory is maintained in the DRAM buffer to support the search operations.) If the page is found in the flash cache (*i.e.*, a flash hit), it is fetched from the flash cache. Otherwise, it is fetched from disk.

- When a page is swapped out of the DRAM buffer, different actions can be taken on the page depending on whether it is clean or dirty. If it is clean, the page is discarded or staged in to the flash cache. If it is dirty, the page is written back to either disk or flash cache, or both.

- When a page is staged out of the flash cache, different actions can be taken on the page depending on whether it is clean or dirty. If it is clean, the page is just discarded. If it is dirty, the page is written to disk unless it was already written to disk when evicted from the DRAM buffer.

Evidently from the key interactions described above, the fundamental issues are when and which data pages should be staged in to or out of the flash cache. There are quite a few alternatives to addressing the issues and they may have profound impact on the overall performance of a database server. For example, when a data page is fetched from disk on a DRAM cache miss, we opt not to enter the page to the flash cache, because the copy in the flash cache will never be accessed while the page is cached in the DRAM buffer. For the reason, staging a page in to the flash cache is considered only when the page is evicted from the DRAM buffer.

### 3.2 Design Choices for *FaCE*

The *FaCE* caching scheme focuses on how exactly data pages are to be staged in the flash cache and how the flash cache should be managed. Specifically, in the rest of this section, we discuss alternative strategies towards the following three key questions and justify the choices made for the *FaCE* scheme: (1) When a dirty page is evicted from the DRAM buffer, does it have to be written through to disk as well as the flash cache or only to the flash cache? (2) Which replacement algorithm is better suited to exploiting flash memory as an extended cache? (3) When a clean page is evicted from the DRAM buffer, does it have to be cached in flash memory at all or given as much preference as a dirty page? These are orthogonal questions. An alternative strategy can be chosen separately for each of the three dimensions.

Note that the *FaCE* scheme does not distinguish data pages by types (*e.g.*, index pages vs. log data) nor monitor data access patterns to separate hot pages from cold ones. Nonetheless, there is nothing in the *FaCE* framework that disallows such additional information to be used in making caching decisions. Table 2 summarizes the design choices of *FaCE* and compares them with existing flash memory caching methods. The design choices of *FaCE* will be elaborated in this section.

|         | Exadata     | TAC         | LC       | *FaCE*   |
|---------|-------------|-------------|----------|----------|
| When    | on entry    | on entry    | on exit  | on exit  |
| What    | clean       | both        | both     | both     |
| Sync    | write-thru  | write-thru  | writeback| writeback|
| Replace | LRU         | Temperature | LRU-2    | FIFO     |

Table 2: Flash Caching Methods

**Write-Back than Write-Through**

When a dirty page is evicted from the DRAM buffer, it may be *written through* to both disk and the flash cache. Alternatively, a dirty page may be *written back* only to the flash cache, leaving its disk copy intact and outdated until it is synchronized with the current copy when being staged out of the flash cache and written to disk.

The two alternative approaches are equivalent in the effect of flash caching for read operations. However, the overhead of the *write-through* policy is obviously much higher than that of the *write-back* policy for write operations. While the *write-through* policy requires a disk write as well as a flash write for each dirty page being evicted from the DRAM buffer, the *write-back* policy reduces disk writes by staging them in the flash cache until they are staged out to disk. In effect, *write-back* replaces one or more disk writes (required for a dirty page evicted repeatedly) with as many flash writes followed by a single (deferred) disk write. For the reason, *FaCE* adopts the *write-back* policy rather than *write-through*.

Now that the flash and disk copies of a data page may not be in full sync, *FaCE* ensures that a request of the page from the database server is always served by the current page. Note that the adoption of a *write-back* policy does not make the database server any more vulnerable to data corruption, as flash memory is a non-volatile storage medium. In fact, the *FaCE* caching scheme provides a low-cost recovery mechanism that takes advantage of the persistency of flash-resident cached data. (Refer to Section 4 for details.)

**Replacement by Multi-Version FIFO**

The size of a flash cache is likely to be much larger than the DRAM buffer but is not unlimited either. The page frames in the flash cache need to be managed carefully for better utilization. The obvious first choice is to manage the flash cache by LRU replacement. With LRU replacement, a victim page is chosen at the rear end of the LRU list regardless of its physical location in the flash cache. This implies that each page replacement incurs a random flash write for an (logically) in-place replacement. Random writes are the pattern that requires flash memory to make the most strenuous effort to process, and the bandwidth of random writes is typically an order of magnitude lower than that of sequential writes with flash memory.

Alternatively, a flash cache can be managed by FIFO replacement. The FIFO replacement may seem irrational given that it is generally considered inferior to the LRU replacement and can suffer from Belady's anomaly with respect to hit rate. Nonetheless, the FIFO replacement has its own unique merit when it is applied to a flash caching scheme. Since all incoming pages are enqueued to the rear end of the flash cache, all flash writes will be done sequentially in the append-only fashion. This particular write pattern is known to be a perfect match with flash memory [13], and helps the flash cache yield the best performance.

The *FaCE* scheme adopts a variant of FIFO replacement. This is different from the traditional FIFO replacement in that one or more different versions of a data page are allowed to be present in the flash cache simultaneously. When a flash frame is to be replaced by an incoming page, a victim frame is selected at the front end of the flash cache and the incoming page is enqueued to the rear end of the flash cache. If the incoming page is dirty, then it is enqueued unconditionally. No additional action is taken to remove existing versions in order not to incur random writes. If the incoming page is clean, then it is enqueued only when the same copy does not exist in the flash cache. A data page dequeued from the flash cache is written to disk only if it is dirty and it is the latest version in the flash cache, so that redundant disk writes can be avoided. We call this approach **Multi-Version FIFO (*mvFIFO*)** replacement. Refer to Section 3.3 for the elaborate design of *FaCE* and its optimization.

Despite the previous studies reporting the limitations of LRU replacement applied to a second level buffer cache [20, 19], we observed that LRU replacement still delivered higher hit rates in the flash cache than the *mvFIFO* replacement. This is partly attributed to the fact that the former keeps no more than a single copy of a page cached in flash memory while the latter may need to store one or more versions of a page. However, the gap in the hit rates was narrow and far outweighed by the reduced I/O cost of the *mvFIFO* replacement with respect to the overall transaction throughput.

**Caching Clean and Dirty**

Caching a page in flash memory will be beneficial, if the cached copy is referenced again before being removed from the flash cache and the cost of a disk read is higher than the combined cost of a flash write and a flash read, which is true for most contemporary disk drives and flash memory SSDs. The caching decision thus needs to be made based on how probable it is a cached page will be referenced again at least once before the page is removed from the flash cache.

As a matter of fact, if a page being evicted from the DRAM buffer is dirty, it is always beneficial to cache the page in flash memory, because an immediate disk write would be requested for the page otherwise. In addition, by caching dirty pages, the *write-back* policy can turn a multitude of disk writes - required for repeated evictions of the same page from the DRAM buffer - into as many flash writes followed by a single disk write. On the other hand, the threshold for caching a clean page in flash memory is higher. This is because caching a clean page could cause a dirty page to be evicted from the flash cache and written to disk, while no disk write would be incurred if the clean page was simply discarded. In that case, the cost of caching a clean page, which could be as high as a disk write and a flash write, will not be recovered by a single flash hit.

Therefore, especially when a *write-back* policy is adopted, dirty pages should have priority over clean ones with respect to caching in flash memory. This justifies the adoption of our new *mvFIFO* replacement that allows multiple versions of a data page to be present in the flash cache.

### 3.3 The *FaCE* System

In a traditional buffer management system, a *dirty* flag is maintained for each page kept in the buffer pool to indicate whether the page has been updated since it was fetched from disk. Using a single flag per page in the buffer is sufficient, because there exist no more than two versions of a data page in the database system at any moment. In the *FaCE* system, however, a data page can reside in the flash cache as well as a DRAM buffer and disk. Therefore, the number of different copies of a page may be more than two, not to mention the different versions of data pages that may be maintained in the flash cache by the *mvFIFO* replacement.

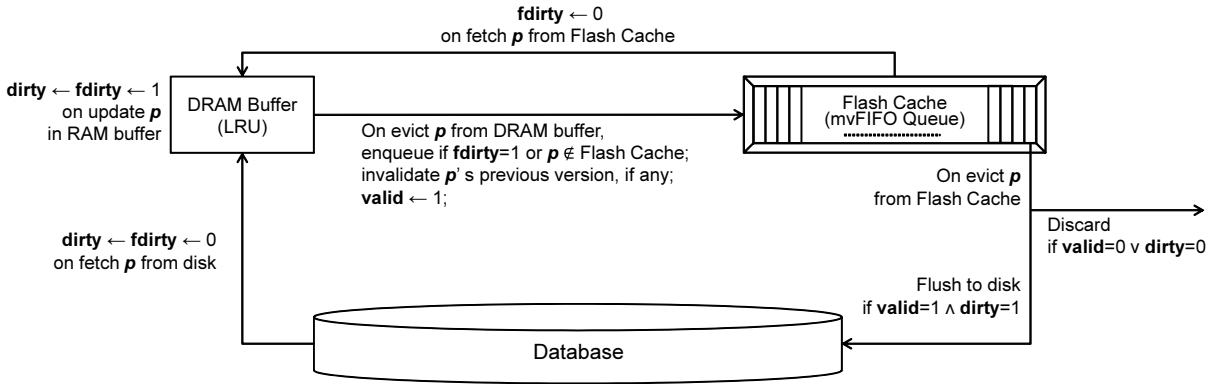

Figure 2: Multi-Version FIFO Flash Cache

We introduce another flag called a *flash dirty* (or *fdirty* in short) to represent the state of a data page accurately. Much like a dirty flag is set on for a data page updated in the DRAM buffer, a *fdirty* flag is set on for a buffered data page that is newer than its corresponding flash resident copy. With the flags playing different roles, the *FaCE* system can determine what action needs to be taken when a page is evicted from the DRAM buffer or from the flash cache. This will be explained in detail shortly.

Both *dirty* and *fdirty* flags are reset to zero, when a page is fetched from disk (because a copy does not exist in the flash cache). They are both set to one when the page is updated in the DRAM buffer. If a page is fetched from the flash cache after being evicted from the DRAM buffer, then the *fdirty* flag must be reset to zero to indicate that the two copies in the DRAM buffer and the flash cache are synced. However, the *dirty* flag of this page must remain unaffected, because the copies in the DRAM and flash cache may still be newer than its disk copy. This requires that while the *fdirty* flags are needed only for the pages in the DRAM buffer, the *dirty* flags must be maintained for the pages both in the DRAM buffer and in the flash cache.

When a page is evicted from the DRAM buffer, it is enqueued to the flash cache unconditionally if the *fdirty* flag is on. Otherwise, it is enqueued to the flash cache only when there is no identical copy already in the flash cache. If a copy is enqueued unconditionally by a raised *fdirty* flag, it will be the most recent version of the page in the flash cache. Since the conditional enqueuing ensures no redundancy in the flash cache, these enqueuing methods – conditional and unconditional – of *mvFIFO* guarantee that there will be no more than a copy of a distinct version and the latest version of a page can be considered the only valid copy among those in the flash cache.

For the convenience of disposing invalid copies, we maintain a *valid* flag for each page cached in the flash memory. The *valid* flag is set on for any page entering the flash cache, which in turn invalidates the previous version in the flash cache so that there exists only a single valid copy at any moment in time. When a page is dequeued from the flash cache, we will flush it to disk only if the *dirty* and *valid* flags are both on. Otherwise, it will be simply discarded. Note that *valid* flags as well as *dirty* and *fdirty* flags are maintained as memory-resident metadata. Invalidating a page copy does not incur any I/O operation. The main operations of the *mvFIFO* replacement are illustrated in Figure 2 and summarized in Algorithm 1.

---
**Algorithm 1**: Multi-Version FIFO Replacement

On eviction of page $p$ from the DRAM buffer:
    **if** $p.fdirty = true \lor p \notin$ *flash cache* **then**
        invalidate the previous version of $p$ if it exists;
        $p$ is enqueued to the flash cache;
        $p.valid \leftarrow true$;
    **endif**
On eviction of page $p$ from the flash cache:
    **if** $p.dirty = true \land p.valid = true$ **then**
        $p$ is written back to disk;
    **else**
        $p$ is discarded;
    **endif**
On fetch of page $p$ from disk:
    $dirty \leftarrow fdirty \leftarrow false$;
On fetch of page $p$ from the flash cache:
    $fdirty \leftarrow false$;
On update of page $p$ in the DRAM buffer:
    $dirty \leftarrow fdirty \leftarrow true$;

---

### Group Second Chance (GSC)

The *Second Chance* replacement is a variant of the FIFO replacement and is generally considered superior to its basic form. The same idea of second chance can be adopted for the *mvFIFO* replacement. With a second chance given to a *valid* page being dequeued from the flash cache, if the page has been referenced while staying in the flash cache, it will be enqueued back instead of being discarded or flushed to disk.

For a DRAM buffer pool, the second chance replacement is implemented by associating a *reference* flag with each page and by having a *clock hand* point to a page being considered for a second chance. Since it does not involve copying or moving pages around, the second chance replacement can be adopted with a little additional cost for setting and resetting *reference* flags. In contrast, for the flash cache being managed by the *mvFIFO* replacement, it is desired that data pages are *physically* dequeued from and enqueued to flash memory for the sake of efficiency of sequential IO operations.

The negative aspect of this, however, is the increased I/O activities, as dequeuing and enqueuing a page require two I/O operations to be made to the flash cache. This will be further aggravated if more than a few valid (and referenced) pages need to be examined by the second chance replacement before a victim page is found in the flash cache. It is an ironic situation, because the more pages in the flash cache are hit by references or utilized, the more likely the cost of a replacement grows higher.

To address this concern, we propose a novel **Group Second Chance (GSC)** replacement for the flash cache. When a page evicted from the DRAM buffer is about to enter the flash cache, a replacement is triggered to make a space for the page. The pages at the front end of the flash cache will be scanned to find a victim page, but the group second chance limits the scan depth so that the replacement cost is bounded. Though the scan depth can be set to any small constant, it will make a practical sense to set the scan depth to no more than the number of pages (typically 64 or 128) in a flash memory block.

All the pages within the scan depth are dequeued from the flash cache in a batch. Following the basic *mvFIFO* replacement, the pages in a batch are either discarded or flushed to disk if their reference flags are down. Then, the remaining pages will be enqueued back to the flash cache. In a rare case where all the pages in the batch are referenced, the page at the very front end will be discarded or flushed to disk in order to make a space for an incoming page from the DRAM buffer. In a more typical case, the number of pages to be enqueued back will be much smaller than the scan depth. In this case, more pages are *pulled* from the LRU tail of the DRAM buffer to fill up the space in the batch. This ensures that a dequeuing or enqueuing operation will be carried out by a single batch-sized I/O operation, much more infrequently than being done for individual pages.

Pulling page frames from the DRAM buffer is analogous to what is done by the background writeback daemons of the Linux kernel or the DBWR processes of the Oracle database server. The Linux writeback daemons wake up when free memory shrinks below a threshold and write dirty pages back to disk to free memory. The Oracle DBWR processes perform a batch write to make clean buffer frames available. We expect that the effect of pulling page frames is negligible on hit rate of the DRAM buffer and the flash cache either positively or negatively.

Note that page replacement in the flash cache can also be done in the batch I/O operations without second chances. We refer to this approach as **Group Replacement (GR)** and compare this with Group Second Chance in Section 5 to better understand the performance impact of the optimizations separately.

## 4. RECOVERY IN *FaCE*

When a system failure happens, it must be recovered to a consistent state such that the atomicity and durability of transactions are ensured. Two fundamental principles for database recovery are write-ahead logging and commit-time force-write of the log tail. The *FaCE* system is no different in that these two principles are strictly applied for database recovery.

When a dirty page is evicted from the DRAM buffer, all of its log records are written to a stable log device in advance. As far as the durability of transactions is concerned, once a dirty page is written to either the flash cache or disk, the page is considered propagated persistently to the database, as the flash memory drive used as a cache extension is non-volatile. The non-volatility of flash memory guarantees that it is always possible for the *FaCE* system to recover the latest copies from the flash cache after a system failure. Therefore, *FaCE* utilizes data pages stored in the flash cache to serve dual purposes, namely, cache extension and recovery.

In this section, we present the recovery mechanism of *FaCE* that achieves transaction atomicity and durability at a nominal cost and recovers the database system from a failure quickly by utilizing data pages survived in the flash cache.

### 4.1 Checkpointing the Flash Cache

If a database system crashes while operating with *FaCE* enabled, there is an additional consideration for the standard restart steps to restore the database to a consistent state. Some of the data pages stored in the flash cache may be newer than the corresponding disk copies, and the database consistency can only be restored by using those flash copies. Even if they were fully synced, the flash copies should be preferred to the disk copies, because the flash copies would help the system restart more quickly.

The only problem in doing this though is that we must guarantee the information about data pages stored in the flash cache survives a system failure. Otherwise, with the information lost, the flash copies of data pages will be inaccessible when the system restarts, and the database may not be restored to a consistent state. Of course, it is not impossible to restore the metadata by analyzing the entire flash cache but it will be an extremely time-consuming process. One practice common in most database systems is to include additional metadata (*e.g.*, file and block identification numbers and `pageLSN`) in the individual page header [9]. However, adding the information about whether it is cached in the flash to the page header is not an option either, because it will incur too much additional disk I/O to update the disk-resident page header whenever a page enters or leaves the flash cache.

One remedy proposed by the temperature-aware caching (TAC) [2] is to maintain the metadata current persistently in the flash memory. The metadata are maintained in a data structure called a slot directory that contains one entry for each data page stored in the flash cache. TAC uses flash memory as a write-through cache and relies on an invalidation-validation mechanism to keep the flash cache consistent with disk at all times. One obvious drawback of this approach is that an entry in the slot directory need be updated for each page entering the flash cache. This overhead will not be trivial, because updating an entry requires two additional random flash writes - one for invalidation and another for validation.

In principle, this burden will be shared by any LRU-based flash caching method that maintains metadata persistently in the flash memory. This is because it needs to update an entry in the metadata directory for each page being replaced in the flash cache, incurring just as much overhead as TAC does. This overhead will be significant and unavoidable, because the LRU replacement selects any page in the flash cache for replacement and, consequently, updating metadata entries will have to be done by random write operations.

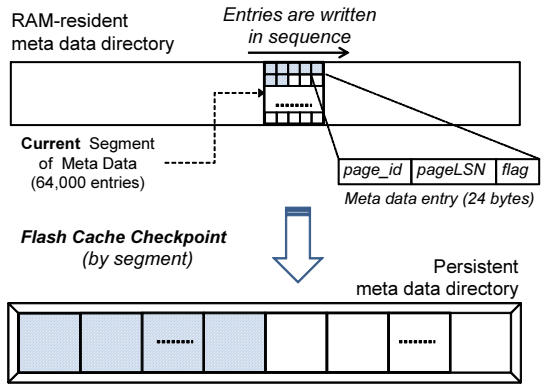

Figure 3: Metadata Checkpointing in *FaCE*

Fortunately, however, *FaCE* has an effective way of dealing with the overhead since it relies on the *mvFIFO* replacement instead of LRU. In a similar way to how a database log tail is maintained, metadata changes are collected in memory and written persistently to flash memory in a single large segment. This type of metadata management is feasible with *FaCE*, because a data page entering the flash cache is always written to the rear in chronological order, and so is its metadata entry to the directory. Therefore, metadata updates will be done more efficiently than doing them for individual entries (typically tens of bytes each). This process is illustrated in Figure 3. We call this metadata flushing operation *flash cache checkpointing*, as saving the metadata persistently (about one and a half MBytes each time in the current implementation of *FaCE*) is analogous to the database checkpointing. The flash cache checkpointing is triggered independently of the database checkpointing.

Of course, the memory resident portion of the metadata directory will be lost upon a system failure, but it can be restored quickly by scanning only a small portion of the flash cache that corresponds to the most recent segment of metadata. Recovering the metadata is described in more detail in the next section.

### 4.2 Database Restart

When a database system restarts after a failure, the first thing to do is to restore the metadata directory of the flash cache. While the vast majority of metadata entries are stored in flash memory and survive a failure, the entries in the current segment are resident in memory and lost upon a failure. To restore the current segment of the metadata directory, the data pages whose metadata entries are lost need be fetched from the flash cache. Those data pages can be found at the rear end of the flash cache maintained as a circular queue. The front and rear pointers are maintained persistently in the flash cache. The data pages in the flash cache contain all the necessary information such as page id and pageLSN in their page header.

In theory, the metadata directory can be restored by fetching the data pages belonging only to the latest segment from the flash cache. However, this will require the database system to be quiesced while flushing the current segment of metadata is in progress, and its negative impact on the performance will not be negligible. In the current implementation of *FaCE*, we allow a new metadata entry to enter the current segment in memory, even when the previous segment is currently being flushed to flash memory. Considering the fact that a failure can happen in the midst of flushing metadata, the current implementation of *FaCE* restores the metadata directory by fetching data pages belonging to the *two most recent segments* of the directory from the flash cache. This way we can avoid quiescing the database system and improve its throughput at minimally increased cost of restart. In fact, as will be shown in Section 5.5, *FaCE* can shorten the overall restart time considerably, because most of the recovery can be carried out by utilizing data pages cached persistently in flash memory.

## 5. PERFORMANCE EVALUATION

We have implemented the *FaCE* system in the PostgreSQL open source database server to demonstrate its effectiveness as a flash cache extension for database workloads. TPC-C benchmark tests were carried out on a hardware platform equipped with a RAID-0 array of enterprise class 15k-RPM disk drives and SLC-type and MLC-type flash memory SSDs.

### 5.1 Prototype Implementation

The *FaCE* system has been implemented as an addition to the buffer manager, recovery and checkpointing modules of PostgreSQL. The most relevant functions in the buffer manager module are `bufferAlloc` and `getFreeBuffer`. The `bufferAlloc` is called upon DRAM buffer misses and it invokes `getFreeBuffer` to get a victim page in the DRAM buffer and to flush it to the database if necessary. We have modified these functions to incorporate the *mvFIFO* and the optimization strategies of *FaCE* in the buffer manager. Specifically, the modified `bufferAlloc` is now responsible for the sync and replacement methods, and the modified `getFreeBuffer` deals with caching clean and/or dirty pages. For database checkpointing, we have modified the `bufferSync` function, which is called from `createCheckPoint`, so that all the dirty pages in the DRAM buffer are checked in to the flash cache instead of disk. A new recovery module for the mapping metadata has been added between the PostgreSQL modules `initBufferPool` and `startupXLOG`.

Besides, we have added a few data structures to manage page frames in the flash cache. Among those are a directory of metadata maintained for all the data pages cached in flash memory and a hash table that maps a page id to a frame in the flash cache. There is an entry for each page cached in the flash memory. Each entry stores its page id, frame id, dirty flag and LSN. Both the metadata directory and the hash table are resident in memory, but a persistent copy of the former is maintained in flash memory as well for recovery purposes. (See Section 4.1 for details.) Since the hash table is not persistent, it is rebuilt from the metadata when the database system restarts.

### 5.2 Experimental Setup

The TPC-C benchmark is a mixture of read-only and update intensive transactions that simulate the activities found in OLTP applications. The benchmark tests were carried out by running PostgreSQL with *FaCE* enabled on a Linux system with 2.8GHz Intel Core i7-860 processor and 4GB DRAM. This computing platform was equipped with an MLC-based SSD (Samsung 470 Series 256GB), an SLC-based SSD (Intel X25-E 32GB), and a RAID-0 disk array with eight drives. The RAID controller was Intel RS2WG160 with PCIe 2.0 interface and 512MB cache, and the disk drives were an enterprise class 15k-RPM Seagate ST3146356SS with 146.8GB capacity each and a Serial Attached SCSI (SAS) interface.

The database size was set to approximately 50GB (scale of 500, approximately 59 GB including indexes and other data files), and the DRAM buffer pool was limited to 200 MB in order to amplify I/O effects for a relatively small database. The capacity of a flash cache was varied between 2GB and 14GB so that the flash cache was larger than the DRAM buffer but smaller than the database. The number of concurrent clients was set to 50, which was the highest level of concurrency achieved on the computing platform before hitting the scalability bottleneck due to contention [10]. The page size of PostgreSQL was 4 KBytes. The benchmark database and workload were created by the BenchmarkSQL tool [14].

For steady-state behaviors, all performance measurements were done after the flash cache was fully populated with data pages. Both the flash memory and disk drives were bound as a raw device, and 'Direct IO' flag was set in PostgreSQL for opening data files so that interference from data caching by the operating system was minimized.

## 5.3 Transaction Throughput

In this section, we analyze the performance impact of *FaCE* with respect to transaction throughput as well as hit rate and I/O utilization. To demonstrate its effectiveness, we compare *FaCE* (and its optimization strategies, GR and GSC, presented in Section 3.3) with the Lazy Cleaning (LC) method [6], which is one of the most recent flash caching strategies based on LRU replacement. Both LC and *FaCE* built into the PostgreSQL database server cache pages – clean or dirty – when they are evicted from the DRAM buffer, and implement the *write-back* policy by copying dirty pages to disk when they are evicted from the flash cache.

**Read Hit Rate and Write Reduction**

Table 3 compares LC and *FaCE* with respect to read hits and write reductions observed in the TPC-C benchmark. As the size of the flash cache increased, both the hit rates and write reductions increased in all cases, because it became increasingly probable for warm pages to be referenced again, overwritten (by LC) or invalidated (by *FaCE*) before they were staged out of the flash cache.

Not surprisingly, the hit rate and write reduction of LC were approximately 2 to 9 percent higher than those of *FaCE* consistently over the entire range of the flash cache sizes tested. This is because, when the flash cache is managed by LC, the flash cache keeps no more than a single copy for each cached page, and the copy is always up-to-date. Thus, LC could utilize the space of the flash cache more effectively than *FaCE*, which could store more than a single copy for each cached page. For example, when the flash cache of 8GB was managed by *FaCE*, the portion of duplicate pages was as high as 30 to 40 percent.

| (Measured in %) | Flash cache size[†] | | | | |
|---|---|---|---|---|---|
| | 2GB | 4GB | 6GB | 8GB | 10GB |
| LC | 72.9 | 80.0 | 83.7 | 87.0 | 89.3 |
| FaCE | 65.5 | 72.6 | 76.4 | 78.6 | 80.5 |
| FaCE+GR | 65.5 | 72.6 | 76.2 | 78.6 | 80.4 |
| FaCE+GSC | 69.7 | 76.6 | 79.8 | 82.1 | 83.7 |

(a) Ratio of flash cache hits to all DRAM misses

| (Measured in %) | Flash cache size[†] | | | | |
|---|---|---|---|---|---|
| | 2GB | 4GB | 6GB | 8GB | 10GB |
| LC | 51.8 | 62.1 | 68.8 | 74.0 | 78.6 |
| FaCE | 46.3 | 54.8 | 60.1 | 62.8 | 65.0 |
| FaCE+GR | 46.3 | 55.3 | 59.7 | 62.7 | 65.4 |
| FaCE+GSC | 50.2 | 59.9 | 65.9 | 70.4 | 73.9 |

(b) Ratio of flash cache writes to all dirty evictions

[†]Database size: 50GB

**Table 3: Read Hit and Write Reduction Rates**

Despite the high ratio of duplicate pages, however, as Table 3 shows, the hit rate of *FaCE* was not lower than that of LC more than 10 percent due to the locality of references in the TPC-C workloads. Another point to note is that the group second chance (GSC) improved not only read hits but also write reductions for *FaCE* by giving dirty pages a second chance to stay and to be invalidated in the flash cache rather than being flushed to disk.

**Utilization and Throughput**

| (Measured in %) | Flash cache size[†] | | | | |
|---|---|---|---|---|---|
| | 2GB | 4GB | 6GB | 8GB | 10GB |
| LC | 92.6 | 96.4 | 97.7 | 98.2 | 98.1 |
| FaCE | 65.6 | 73.7 | 78.9 | 82.7 | 84.9 |
| FaCE+GR | 51.6 | 62.5 | 67.7 | 70.0 | 69.6 |
| FaCE+GSC | 60.9 | 68.0 | 70.9 | 74.7 | 75.9 |

(a) Device-level utilization of the flash cache

| (Measured in IOPS) | Flash cache size[†] | | | | |
|---|---|---|---|---|---|
| | 2GB | 4GB | 6GB | 8GB | 10GB |
| LC | 4534 | 4226 | 3849 | 3362 | 3370 |
| FaCE | 4973 | 5870 | 6479 | 7019 | 7415 |
| FaCE+GR | 7213 | 8474 | 9390 | 9848 | 10693 |
| FaCE+GSC | 11098 | 12208 | 13031 | 13871 | 14678 |

(b) Throughput of 4KB-page I/O operations

[†]Database size: 50GB

**Table 4: Utilization and I/O Throughput**

Keeping a single copy for each cached page by LRU often requires overwriting an existing copy – either an old version or a victim page – in the flash cache. Overwriting a page in the flash cache will most likely incur a random flash write. Table 4 compares LC and *FaCE* with respect to average utilization and I/O throughput of the flash cache. Table 4(a) shows that LC raised the device saturation of a flash cache quickly to 92 percent or higher. The flash cache device was never saturated more than 76 percent by *FaCE* with the GSC optimization.

Table 4(b) compares LC and *FaCE* with respect to throughput measured in 4KB page I/O operations carried out per second. The results clearly reflects the difference between LC and *FaCE* in the device saturation of the flash cache. *FaCE* with group second chance (GSC) processed I/O operations more efficiently than LC by more than a factor of four when the size of the flash cache was 10GB. This was because the write operations were dominantly random by LC while they were dominantly sequential by *FaCE*. More importantly, as the size of the flash cache increased, the I/O throughput of *FaCE* improved consistently and considerably while that of LC deteriorated. This was caused by a common trend of writes that the randomness becomes higher as the data region of writes is extended [12].

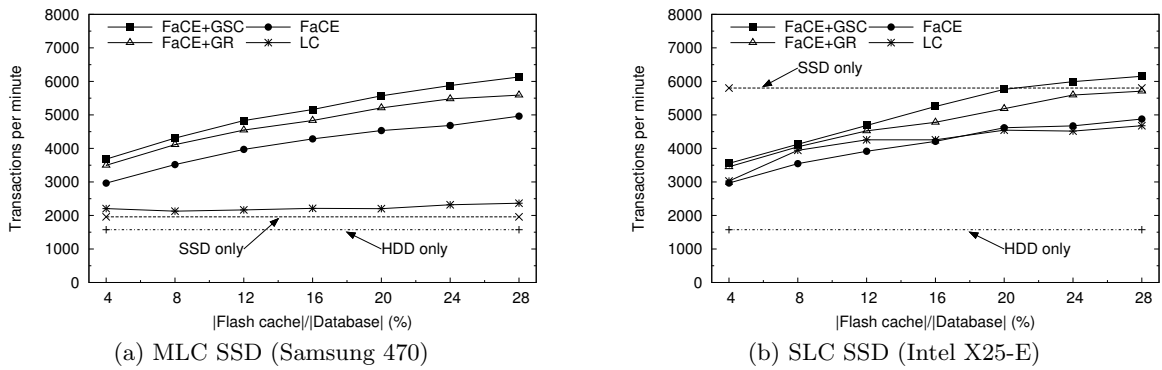

Figure 4: Transaction Throughput: LC vs. *FaCE*

**Impacts on Transaction Throughput**

Figure 4 compares LC and *FaCE* with respect to transaction throughput measured in the number of transactions processed per minute (tpmC). In order to analyze the effect of different types of flash memory SSDs, we used both MLC-type (Samsung 470) and SLC-type (Intel X25-E) SSDs in the experiments. (Refer to Table 1 for the characteristics of the SSDs.) Besides, in order to understand the scope of performance impact by flash caching, we included the cases where the database was stored entirely on either a flash memory SSD (denoted by `SSD-only`) or a disk array (denoted by `HDD-only`).[2]

Figure 4(a) shows the transaction throughput obtained by using an MLC SSD as a flash caching device (or as a main storage medium for the `SSD-only` case). Under the LC method, the transaction throughput remained flat without any significant improvements with increases in the size of flash cache. This is because the MLC SSD was already utilized at its maximum (93 to 98 percent as shown in Table 4), and the saturation in the flash caching device became a performance bottleneck quickly as more data pages were stored in the flash cache. Under the *FaCE* method, in contrast, the utilization of flash caching device was below 85 percent consistently. As the size of the flash cache increased, the transaction throughput continued to improve over the entire range without being limited by the I/O throughput of the flash caching drive.

With an SLC SSD, as is shown in Figure 4(b), almost identical trend was observed in transaction throughput by *FaCE*.

The LC method, on the other hand, improved transaction throughput and reduced the performance gap considerably. This was due to the higher random write throughput of the SLC SSD. With the group second chance optimization, however, *FaCE* still outperformed LC at least 25 percent. As is pointed out in Section 2, the bandwidth disparity between random writes and sequential writes still exists even in the SLC type of flash memory SSDs, and *FaCE* was apt to take advantage of it better by turning random writes into sequential ones.

For the same reason, by utilizing a small flash cache, *FaCE* achieved higher transaction throughput even than `SSD-only`, which would have to deal with a considerable amount of random writes. In the case of an MLC SSD, *FaCE* outperformed `SSD-only` almost by three folds. This striking result demonstrates the cost-effectiveness of *FaCE* in terms of return on investment.

In summary, while LC achieved hit rates and reduced the amount of traffic to disk drives better, *FaCE* was superior to LC in utilizing the flash cache efficiently for higher I/O throughput. Despite the trade-off, it turned out the benefit from higher I/O bandwidth of flash cache outweighed the benefit from reduced disk I/O operations. Overall, *FaCE* outperformed LC considerably in transaction throughput irrespective of the type of a flash memory device used.

## 5.4 Cost Effectiveness and Scalability

This section evaluates empirically the cost effectiveness of the flash cache and its impact on the scalable throughput of a disk array. These two aspects of flash caching have not been studied in the existing work.

### 5.4.1 More DRAM or More Flash

A database buffer pool is used to replace slow disk accesses with faster DRAM accesses for frequently accessed database items. In general, investing resources in the DRAM buffer is an effective means of improving the overall performance of a database system, because a larger DRAM buffer pool will reduce page faults and increase the throughput. As the size of a DRAM buffer increases, however, the return on investment will not be sustained indefinitely and will eventually saturate after passing a certain threshold. For example, with TPC-C workloads, the DRAM buffer hit rate is known to reach a knee point when its size is quite a small fraction of the database size due to skewed data accesses [18].

As the analysis given in Section 2.2 indicates, the return on investment is likely to be much higher with flash memory than DRAM given the persistent and widening price gap between them. To demonstrate the cost effectiveness of flash memory as a cache extension, we measured the performance gain that would be obtained by making the same amount of monetary investment to DRAM and flash memory.

---

[2] For the `SSD-only` with an SLC SSD, the database size had to be reduced 20 percent – 400 warehouses instead of 500 – due to the limited capacity. 64GB was the largest capacity of X25-E at the time of experiments and was not large enough to store a 50GB database, its indexes and other related data.

| (Measured in tpmC) | 200MB DRAM or 2GB Flash | | | | |
|---|---|---|---|---|---|
| | x1 | x2 | x3 | x4 | x5 |
| More DRAM | 2061 | 2353 | 2501 | 2705 | 2843 |
| More Flash | 3681 | 4310 | 4830 | 5161 | 5570 |

**Table 5: More DRAM vs. More Flash**

Assuming that the cost per gigabyte of DRAM is approximately ten times higher than that of MLC-type flash memory [7], we evaluated the cost effectiveness by measuring the throughput increment obtained from each 2GB of flash memory or alternatively each 200MB of DRAM added to the basic system configuration described in Section 5.2. As is shown in Table 5, the transaction throughput (or return on investment) was consistently higher with a wide margin when more resources were directed to flash memory rather than DRAM. Flash caching was disabled in the first row of the table, and *FaCE* +GSC was used in the second row.

### 5.4.2 Scale-Up with More Disks

No matter how large a DRAM buffer or a flash cache is, there will be cache misses if neither is as large as the entire database. For the reason, the I/O throughput of disk drives will always remain on the critical path of most database operations. For example, when a data page is about to enter the flash cache, another page may have to be staged out of the flash cache if it is already full. In this case, a page evicted from the DRAM buffer will end up causing a disk write followed by a flash write. This acutely points out the fact that the system throughput may not be improved up to its full potential by the flash cache alone without addressing the bottleneck (*i.e.*, disks) on the critical path.

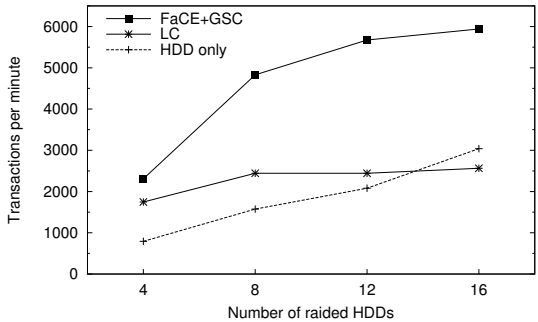

**Figure 5: Effect of a Disk Array Size**

Using a disk array is a popular way of increasing disk throughput (measured in IOPS). Figure 5 compares LC and *FaCE* with respect to transaction throughput measured with a varying number of disk drives from four to sixteen. (For this test, we added eight more disk drives to the basic system configuration described in Section 5.2.) The flash cache size was set to 6GB for both LC and *FaCE*, and `HDD-only` was also included in the scalability comparison. The same database was distributed across all available disk drives.

For *FaCE* and `HDD-only`, transaction throughput increased consistently with the increase in the number of disk drives.

This confirms that disk drives were on the critical path and increasing disk throughput was among the keys to improving the transaction throughput. It also confirms that, with *FaCE*, the flash cache was never a limiting factor in transaction throughput until the number of disk drives was increased to sixteen, at which point the transaction throughput started saturating. In contrast, the transaction throughput of LC did not scale and improve at all beyond eight disk drives, and became even worse than that of `HDD-only` when sixteen disk drives were used.

## 5.5 Performance of Recovery

In order to evaluate the recovery system of *FaCE*, we crashed the PostgreSQL server using the Linux `kill` command and measured the time taken to restart the system. When database checkpointing was turned on, the `kill` command was issued at the mid-point of a checkpoint interval. For example, if the checkpoint interval was 180 seconds, the `kill` command was issued 90 seconds after the most recent checkpoint. In the case of *FaCE*, the GSC optimization was enabled for the flash cache management. The flash cache size was set to 4GB.

| (Measured in second) | Checkpoint intervals | | |
|---|---|---|---|
| | 60 | 120 | 180 |
| FaCE+GSC | 93 | 118 | 188 |
| HDD only | 604 | 786 | 823 |

**Table 6: Time Taken to Restart the System**

Table 6 presents the average recovery time taken by the system when it was run with different checkpoint intervals 60, 120, and 180 seconds. For each checkpoint interval, we took the average of restart times measured from five separate runs. Across all three checkpoint intervals, *FaCE* reduced the restart time considerably – from 77 to 85 percent – over the system without flash caching. Such significant reduction in restart time was possible because the recovery could be carried out by utilizing data pages cached persistently in flash memory. We observed in our experiments that more than 98 percent of data pages required for recovery were fetched from the flash cache instead of disk.

The restart times given in Table 6 for *FaCE* include the time taken to restore the metadata directory, which was approximately 2.5 seconds on average regardless of the checkpoint interval. The metadata directory can be restored by fetching the persistent portion of the directory from flash memory and by scanning as many data pages as two segments worth of metadata entries from the flash cache. The latter is required to rebuild the most recent segment of the directory. In our experiments, the persistent portion of the metadata directory was 21 MBytes and the amount of data pages to read from the flash cache was 512 MBytes.[3] The sequential read bandwidth of the flash memory SSD used in our experiments was high enough – at least 250 MB/sec – to finish this task within 2.5 seconds on average.

Figure 6 shows the time-varying transaction throughput measured immediately after the system was recovered from a failure. This figure clearly demonstrates that, when *FaCE*

---
[3]Each segment contains 64,000 metadata entries of 24 Bytes each. Among the 16 segments required for a 4GB flash cache, 14 of them are fetched directly from flash memory and the rest are rebuilt from the data page headers.

1625

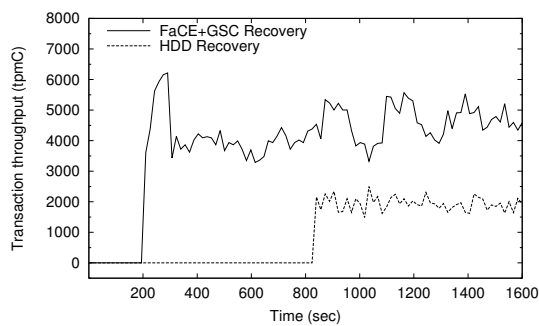

Figure 6: Transaction Throughput after Restart

was enabled, the system resumes normal transaction processing much more quickly and maintains higher transaction throughput at all times. The checkpoint interval was set to 180 seconds in this experiment.

## 6. CONCLUSION

This paper presents a low-overhead caching method called *FaCE* that utilizes flash memory as an extension to a DRAM buffer for a recoverable database. *FaCE* caches data pages in flash memory on exit from the DRAM buffer. By basing its caching decision solely on the DRAM buffer replacement, the flash cache is capable of sustaining high hit rates without incurring excessive run-time overheads for monitoring access patterns, identifying hot and cold data items, and migrating them between flash memory and disk drives. We have implemented *FaCE* and its optimization strategies within the PostgreSQL open source database server, and demonstrated that *FaCE* achieves a significant improvement in the transaction throughput.

We have also made a few important observations about the effectiveness of *FaCE* as a flash caching method. First, *FaCE* demonstrates that adding flash memory as a cache extension is more cost effective than increasing the size of a DRAM buffer. Second, the optimization strategies (*i.e.*, GR and GSC) of *FaCE* indicate that turning small random writes to large sequential ones is critical to maximizing the I/O throughput of a flash caching device so as to achieve scalable transaction throughput. Third, the *mvFIFO* replacement of *FaCE* enables efficient and persistent management of the metadata directory for the flash cache, and allows more sustainable I/O performance for higher transaction throughput. Fourth, *FaCE* takes advantage of the non-volatility of flash memory to minimize the recovery overhead and accelerate the system restart from a failure. Since most data pages needed during the recovery phase tend to be found in the flash cache, the recovery time can be shortened significantly.

## Acknowledgments

This work was supported by National Research Foundation of Korea (NRF) funded by the Ministry of Education, Science and Technology (No. 2011-0026492, No. 2011-0027613), and the IT R&D program of MKE/KEIT (10041244, SmartTV 2.0 Software Platform).